\documentclass[aps,prb,reprint,floatfix,superscriptaddress]{revtex4-2}
\usepackage{amsmath,amssymb,amsthm}
\usepackage{physics}
\usepackage[dvipdfmx]{graphicx}
\usepackage{appendix}
\usepackage[bookmarks=false,colorlinks=true,linkcolor=blue,filecolor=blue,citecolor=blue,urlcolor=blue]{hyperref}
\usepackage[dvipsnames]{xcolor}
\usepackage{mathrsfs}
\usepackage{bm}
\usepackage[normalem]{ulem}

\begin{document}

\title{Effective Hamiltonian description on monitored Majorana chains: correlated power-law hoppings and unconventional entanglement scaling}

\author{Ken Mochizuki}
\affiliation{Department of Applied Physics, University of Tokyo, 7-3-1 Hongo, Bunkyo-ku, Tokyo 113-8656, Japan}
\affiliation{Nonequilibrium Quantum Statistical Mechanics RIKEN Hakubi Research Team, RIKEN Pioneering Research Institute (PRI), 2-1 Hirosawa, Wako, Saitama 351-0198, Japan}

\author{Hisanori Oshima}
\affiliation{Department of Physics, University of Warwick, Coventry, CV4 7AL, United Kingdom}

\author{Ryusuke Hamazaki}
\affiliation{Nonequilibrium Quantum Statistical Mechanics RIKEN Hakubi Research Team, RIKEN Pioneering Research Institute (PRI), 2-1 Hirosawa, Wako, Saitama 351-0198, Japan}
\affiliation{RIKEN Center for Interdisciplinary Theoretical and Mathematical Sciences (iTHEMS), RIKEN, Wako 351-0198, Japan}

\author{Yohei Fuji}
\affiliation{Department of Physical Sciences, Ritsumeikan  University, 1-1-1 Noji-higashi, Kusatsu, Shiga 525-8577, Japan}

\date{\today}

\begin{abstract}
We investigate the structure of effective Hamiltonians governing monitored dynamics of a one-dimensional Majorana chain using Lyapunov spectral analysis. 
We focus on the gapless phase of the Lyapunov spectrum, where the effective Hamiltonians behave differently from conventional critical Hamiltonians in isolated systems; the spectral gap closes faster than $1/L$ but slower than $1/L^2$ and the entanglement entropy grows as $[\ln(L)]^2$ where $L$ is the system size. We find that the effective Hamiltonians have random long-range power-law hoppings with nontrivial magnitude correlations, rather than being independently and identically distributed. To elucidate the role of these non-Gaussian correlations, we construct random power-law hopping models that capture the essential features of the effective Hamiltonians. The spectral gaps of the constructed models decay faster than $1/L$ but slower than $1/L^2$. We find that, in the absence of hopping correlations, the ground-state entanglement exhibits $\ln(L)$ scaling. In the presence of correlations, by contrast, the entanglement entropy is enhanced and its system-size dependence is consistent with $[\ln(L)]^2$ scaling over the system sizes studied. These results suggest that correlations among long-range hopping magnitudes are responsible for the entanglement scaling rarely encountered in the ground states of conventional isolated quantum systems. 
\end{abstract}

\maketitle

\section{Introduction}
\label{sec:introduction}

Driven by the rapid development of quantum hardware that directly operates many-body quantum states, monitored quantum systems have emerged as a new arena for exploring nonequilibrium quantum physics. Prominent examples of intriguing phenomena are the measurement-induced phase transitions, nontrivial purification dynamics, and unconventional entanglement scalings~\cite{li2018quantum,cao2019entanglement,chan2019unitary,li2019measurement,skinner2019measurement,szyniszewski2019entanglement,bao2020theory,choi2020quantum,fuji2020measurement,gullans2020dynamical,lunt2020measurement,szyniszewski2020universality,turkeshi2020measurement,alberton2021entanglement,lu2021spacetime,agrawal2022entanglement,barratt2022field,block2022measurement,jian2022criticality,minato2022fate,muller2022measurement,noel2022measurement,sierant2022universal,zabalo2022operator,fava2023nonlinear,granet2023volume,kells2023topological,klocke2023majorana,koh2023measurement,le2023volume,loio2023purification,majidy2023critical,mochizuki2023distinguishability,oshima2023charge,poboiko2023theory,yamamoto2023localization,aziz2024critical,chakraborty2024charge,fava2024monitored,kumar2024boundary,le2024entanglement,deluca2025universality,klocke2025entanglement,mochizuki2025measurement,mochizuki2025transitions,oshima2025topology,poboiko2025measurement,bhuiyan2026free,fan2026entanglement,guo2026measurement,guo2026super,hamazaki2026an,poboiko2026quantum,xiao2026symmetry}. Because these are properties of individual quantum trajectories rather than of the averaged density matrix, they have no direct counterpart in the equilibrium phases of isolated quantum systems.

To characterize these transitions and purification dynamics, effective Hamiltonians governing the long-time behavior of monitored quantum systems have recently been introduced through Lyapunov spectral analysis~\cite{zabalo2022operator,aziz2024critical,bulchandani2024random,chakraborty2024charge,kumar2024boundary,deluca2025universality,mochizuki2025measurement,oshima2025topology,mochizuki2025transitions,yokomizo2025measurement,hamazaki2026an}. In this framework, a quantum trajectory generated by long-time monitored dynamics is identified with the ground states of effective Hamiltonians whose energy spectrum is given by the Lyapunov exponents, so that the transitions and purification dynamics can be discussed in terms of the spectral gap and the ground-state entanglement. This description offers a natural bridge between monitored and isolated quantum systems. However, the effective Hamiltonians obtained in this way differ from the Hamiltonians conventionally studied in isolated systems in two respects: they are inherently random, inheriting the stochastic nature of the measurement outcomes, and nothing guarantees that their couplings are short-ranged.

This is a crucial issue, since the range of the couplings largely determines the universal properties of ground states. For short-range Hamiltonians, a variety of such properties have been established, including the stability and the entanglement area law of gapped ground states~\cite{lieb1972finite,hastings2005quasiadiabatic,hastings2006spectral,hastings2007area,bravyi2010topological,chen2010local,brandao2015exponential,arad2017rigorous,sachdev2001quantum,zeng2019quantum}. Long-range Hamiltonians, in which interactions or hoppings decay algebraically with distance, are instead known to exhibit phenomena that are absent in short-range systems~\cite{campa2009statistical,defenu2023long,chen2023speed}, such as Anderson transitions in one dimension~\cite{levitov1989absence,levitov1990localization,mirlin1996transition,evers2000fluctuation,mirlin2000multifractality,varga2000critical,lima2004finite,evers2008anderson,pouranvari2014maximally,roy2018entanglement,zhang2024magnetic,pain2026robust} and violations of the entanglement area law~\cite{koffel2012entanglement,pouranvari2014maximally,vodola2014kitaev,vodola2016long,roy2019quantum,juhasz2022testing,solfanelli2023logarithmic,chakraborty2024entanglement,pereira2026long}. If the effective Hamiltonians of monitored systems are long-ranged, their ground states may inherit this phenomenology. However, whether this is the case, and which features of the effective Hamiltonians are responsible for the unconventional behavior found in monitored dynamics, remain to be clarified.

In this paper, we investigate the structure of effective Hamiltonians describing monitored Majorana chains, focusing on the gapless phase of the Lyapunov spectrum. We find that the corresponding effective Hamiltonians contain random long-range hoppings whose magnitudes decay algebraically with distance. Importantly, the ensemble averages of these hoppings themselves vanish, implying that the effective Hamiltonians cannot be understood from translationally invariant Hamiltonians with weak disorder. Moreover, the hoppings are not independently and identically distributed but instead exhibit nontrivial correlations. We also show that the ground-state entanglement scales as $[\ln(L)]^2$ and that the spectral gap decays faster than $1/L$ but slower than $1/L^2$; the former feature is absent in the uncorrelated random power-law hopping model with the same decay exponent. To clarify the role of the non-Gaussian hopping correlations in these unconventional scalings, we further construct random Hamiltonians with power-law hoppings in which the strength of the correlations between hopping magnitudes can be controlled. We numerically show that the gap of the constructed model also decays faster than $1/L$ but slower than $1/L^2$. We also find that the controlled correlations among hopping magnitudes change the ground-state entanglement scaling from $\ln(L)$ to $[\ln(L)]^2$ over system sizes explored here. These results suggest that long-range hoppings and correlated randomness in effective Hamiltonians are essential for explaining unconventional entanglement scalings observed in monitored quantum systems.

The rest of this paper is organized as follows. In Sec.~\ref{sec:monitored-majorana}, after reviewing the dynamics of monitored Majorana chains, we examine the finite-size scaling of the spectral gap and the ground-state entanglement entropy of the corresponding effective Hamiltonians and identify nontrivial correlations among their hopping magnitudes. In Sec.~\ref{sec:long-range-correlated_Hamiltonians}, we construct random long-range hopping Hamiltonians with non-Gaussian correlations and investigate the scaling of their spectral gaps and ground-state entanglement entropy. Section~\ref{sec:conclusion} concludes the paper.

\section{Monitored Majorana circuits}
\label{sec:monitored-majorana}
\begin{figure}[tbp]
\begin{center}
\includegraphics[width=6cm]{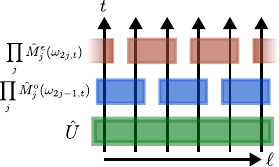}
\caption{The schematic picture of the monitored Majorana circuit. The red and blue rectangles correspond to Majorana parity measurements on even and odd bonds, respectively. The green rectangle represents the unitary dynamics by $\hat{U}$.} 
\label{fig:circuit}
\end{center}
\end{figure}

We consider $2L$ Majorana fermions on a one-dimensional chain, described by Hermitian Majorana operators $\hat{\gamma}_\ell=\hat{\gamma}_\ell^\dagger$ satisfying $\{\hat{\gamma}_\ell,\hat{\gamma}_{\ell'}\}=2\delta_{\ell\ell'}$. The system evolves through alternating unitary evolution and parity measurements of nearest-neighbor Majorana pairs~\cite{oshima2025topology}, as illustrated in Fig.~\ref{fig:circuit}. The unitary evolution is described by
\begin{align}
    \hat{U} = \exp\left(J \sum_{\ell=1}^{2L} \hat{\gamma}_{\ell}\hat{\gamma}_{\ell+1}\right),
    \label{eq:unitary-operator}
\end{align}
with $J$ being a real number. When the measurements of Majorana pairs on the odd bonds $(2j-1,2j)$ and even bonds $(2j,2j+1)$ are performed, the quantum state is transformed by the Kraus operators defined as
\begin{align}
\hat{M}^o_j(\omega)&=\frac{1}{\sqrt{2\cosh (2\theta_o)}}\exp\left(-i\omega\theta_o \hat{\gamma}_{2j-1} \hat{\gamma}_{2j}\right),
\label{eq:Kraus-operator_odd}\\
\hat{M}^e_j(\omega)&=\frac{1}{\sqrt{2\cosh (2\theta_e)}}\exp\left(-i\omega\theta_e \hat{\gamma}_{2j} \hat{\gamma}_{2j+1}\right),
\label{eq:Kraus-operator_even}
\end{align}
respectively, where $\omega=\pm1$ is the corresponding measurement outcome. In the following discussion, we characterize the measurement strength by $\mu_{o/e} = \tanh(\theta_{o/e})$, which corresponds to a projective measurement for $\mu_{o/e}=1$ and to the identity operation for $\mu_{o/e}=0$. We impose the periodic boundary condition $\hat{\gamma}_{2L+1}=\hat{\gamma}_1$ in both $\hat{U}$ and $\hat{M}^e_L(\omega)$. The quantum state at the time step $t$, denoted by $\ket{\psi(\bm{\omega}_t)}$, is conditioned on the sequence of measurement outcomes obtained from the first to $t$ steps. We denote the full measurement record up to time $t$ by $\bm{\omega}_t=\{ \omega_1, \ldots, \omega_t \}$, where $\omega_t$ denotes the outcomes of all measurements performed at the step $t$, $\omega_t = \{ \omega_{1,t}, \ldots, \omega_{2L,t} \}$. For a given measurement record $\bm{\omega}_t$, the time evolution at $t$ becomes
\begin{align}
    \ket{\psi(\bm{\omega}_t)}=\frac{\hat{M}(\omega_t)\ket{\psi(\bm{\omega}_{t-1})}}{\sqrt{p(\omega_t|\bm{\omega}_{t-1})}},
    \label{eq:time-evolution_one-step_many-body}
\end{align}
where $\hat{M}(\omega_t)$ is the one-step time-evolution operator,
\begin{align}
    \hat{M}(\omega_t) = 
    \left[\prod_{j}\hat{M}_j^e(\omega_{2j,t})\right]
    \left[\prod_{j}\hat{M}_j^o(\omega_{2j-1,t})\right]
    \hat{U}.
    \label{eq:Kraus-operator_one-step_many-body}
\end{align}
Here, $p(\omega_t|\bm{\omega}_{t-1})=\bra{\psi(\bm{\omega}_{t-1})}\hat{M}^\dagger(\omega_t)\hat{M}(\omega_t)\ket{\psi(\bm{\omega}_{t-1})}$ is the Born probability of obtaining the set of outcomes $\omega_t$, conditioned on the previous measurement record $\bm{\omega}_{t-1}$. The dynamics conserves the total Majorana parity $\hat{P}=(-i)^L\prod_\ell\hat{\gamma}_\ell$: $[\hat{P},\hat{M}(\omega_t)]=0$. Throughout this paper, we take the initial state as the vacuum and focus on the even-parity sector.

\subsection{Effective Hamiltonians}
\label{subsec:effective-Hamiltonian}
We analyze the dynamics of monitored Majorana fermions by introducing an effective Hamiltonian conditioned on the measurement record,
\begin{align} \label{eq:effective-hamiltonian}
    \hat{\mathcal{K}}(\bm{\omega}_t)=-\frac{1}{2t}\ln\left[\hat{\mathcal{M}}(\bm{\omega}_t)\hat{\mathcal{M}}^\dagger(\bm{\omega}_t)\right],
\end{align}
based on Lyapunov analysis~\cite{mochizuki2025measurement,oshima2025topology,hamazaki2026an}. Here, $\hat{\mathcal{M}}(\bm{\omega}_t)$ is the $t$-step time-evolution operator composed of the Kraus operators and the unitary operators,
\begin{align}
    \hat{\mathcal{M}}(\bm{\omega}_t)=\hat{M}(\omega_t)\hat{M}(\omega_{t-1})\cdots\hat{M}(\omega_1),
    \label{eq:Kraus-operator_t-step_many-body}
\end{align}
where $\hat{M}(\omega_t)$ is defined in Eq.~\eqref{eq:Kraus-operator_one-step_many-body}. In the long-time limit, $\ket{\psi(\bm{\omega}_t)}$ becomes the ground state of the effective Hamiltonian $\hat{\mathcal{K}}(\bm{\omega}_t)$. This correspondence allows us to characterize measurement-induced phases in monitored dynamics in terms of quantum phases of the effective Hamiltonian $\hat{\mathcal{K}}(\bm{\omega}_t)$.

Since the generators of $\hat{U}$ and $\hat{M}_j^{o/e}(\omega)$ are quadratic in $\{\hat{\gamma}_\ell\}$, $\hat{\mathcal{K}}(\bm{\omega}_t)$ can also be written in the quadratic form,
\begin{align} \label{eq:effham-single-to-many}
    \hat{\mathcal{K}}(\bm{\omega}_t)=-\frac{i}{4}\Vec{\gamma}^\mathsf{T} \mathcal{K}(\bm{\omega}_t) \Vec{\gamma} + c\hat{\mathbb{I}},
\end{align}
where $\hat{\mathbb{I}}$ is the identity operator and $c$ is a constant independent of the measurement record $\bm{\omega}_t$, whose explicit form is given in Eq.~\eqref{eq:constant_shift}.
Here, we have defined $\Vec{\gamma}=(\hat{\gamma}_1,\ldots,\hat{\gamma}_{2L})^\mathsf{T}$. The single-particle effective Hamiltonian, which is a $2L \times 2L$ real anti-symmetric matrix $\mathcal{K}(\bm{\omega}_t)=\mathcal{K}^*(\bm{\omega}_t)=-\mathcal{K}^\mathsf{T}(\bm{\omega}_t)$,  becomes
\begin{align}
    \mathcal{K}(\bm{\omega}_t) = \frac{i}{2t}\ln[\mathcal{M}(\bm{\omega}_t)\mathcal{M}^\dagger(\bm{\omega}_t)],
    \label{eq:effective-Hamiltonain_single-particle}
\end{align}
where $\mathcal{M}(\bm{\omega}_t)$ is defined by~\cite{oshima2025topology}
\begin{align}
    [\hat{\mathcal{M}}(\bm{\omega}_t)]^\dagger\Vec{\gamma}[\hat{\mathcal{M}}^{-1}(\bm{\omega}_t)]^\dagger=\mathcal{M}(\bm{\omega}_t)\Vec{\gamma}.
\end{align}
The effective Hamiltonians $\mathcal{K}(\bm{\omega}_t)$ belong to the symmetry class D (BDI) in the presence (absence) of $\hat{U}$, as shown in Appendix~\ref{sec:symmetry}. In our numerical simulations, we evolve the correlation matrix using a procedure based on QR decomposition. Further details on the numerical simulation of monitored dynamics and the construction of effective Hamiltonians are provided in the Supplemental Material of Ref.~\cite{oshima2025topology}. We note that the average of the effective Hamiltonian $\mathcal{K}(\bm{\omega}_t)$ vanishes, as shown in Appendix~\ref{sec:Hamiltonian-average}. Thus, unlike conventional weakly disordered systems, $\mathcal{K}(\bm{\omega}_t)$ cannot be viewed as a translationally invariant Hamiltonian perturbed by small random terms.

We consider the eigenvalues of the single-particle effective Hamiltonian $\mathcal{K}(\bm{\omega}_t)$, which we denote by $\{\pm i\varepsilon_1(\bm{\omega}_t),\ldots,\pm i\varepsilon_L(\bm{\omega}_t)\}$. Here we arrange them in descending order, $\varepsilon_\nu(\bm{\omega}_t)\geq\varepsilon_{\nu+1}(\bm{\omega}_t)\geq0$. For sufficiently large $t$, these eigenvalues converge to constants $\{\pm i\varepsilon_1,\ldots,\pm i\varepsilon_L\}$ that are independent of the measurement outcomes $\bm{\omega}_t$, which we call the single-particle Lyapunov spectrum:
\begin{align}
    \lim_{t\rightarrow\infty}\varepsilon_\nu(\bm{\omega}_t)=\varepsilon_\nu.
\end{align}
This convergence originates from the fact that the averaged CPTP dynamics has a unique and positive-definite stationary state in each parity sector, as shown in Appendix~\ref{sec:ergodicity}~\cite{benoist2019invariant,hamazaki2026an}. In our numerical simulations, we follow a single quantum trajectory and regard $\varepsilon_\nu(\bm{\omega}_t)$ as converged when 
\begin{align}\overline{[\varepsilon_\nu(\bm{\omega}_t)-\overline{\varepsilon_\nu(\bm{\omega}_t)}]^2}\leq\left[w\overline{\varepsilon_\nu(\bm{\omega}_t)}\right]^2
\label{eq:convergence-condition}
\end{align}
is satisfied for sufficiently small $w$. Here and in the rest of this section, we denote by $\overline{X(\bm{\omega}_t)}$ the time average of a trajectory quantity $X(\bm{\omega}_t)$ over $N_s$ consecutive time steps, from $t-N_s+1$ to $t$. 
As shown in Appendix~\ref{sec:spectral-width}, the spectral width is bounded by a constant independent of the system size $L$, $2\varepsilon_1\leq4(\theta_o+\theta_e)$. We have also numerically confirmed that $\sum_{\ell m}\mathcal{K}^2_{\ell m}(\bm{\omega}_t)/L=2\sum_\nu\varepsilon_\nu^2(\bm{\omega}_t)/L=O(1)$ is satisfied for any $L$ studied here, while the data are not shown.

\begin{figure}[tbp]
\begin{center}
\includegraphics[width=6cm]{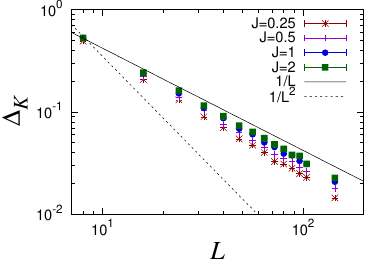}
\caption{The spectral gaps of the effective Hamiltonians for the monitored Majorana circuits in various system sizes, with $\mu_o=\mu_e=0.5$, $w=10^{-3}$, and $N_s=10^3$. Green squares, blue circles, purple crosses, and red asterisks correspond to $J=2$, $1$, $0.5$, and $0.25$, respectively. The black solid and dashed lines are proportional to $1/L$ and $1/L^2$, respectively. } 
\label{fig:gap_oshima-model}
\end{center}
\end{figure}

We define the finite-size spectral gap as the spacing between the two single-particle Lyapunov levels closest to zero,
\begin{align}
    \Delta_K=2\varepsilon_L.
\end{align}
As established in Ref.~\cite{oshima2025topology}, the monitored Majorana chain exhibits three phases: trivial and topological gapped phases and a gapless phase. In both gapped phases, $\Delta_K$ remains finite in the thermodynamic limit, and these phases share similar properties with the gapped ground states of the conventional Kitaev chain. By contrast, the gapless phase exhibits an unusual finite-size scaling: as shown in Fig.~\ref{fig:gap_oshima-model}, the spectral gap closes faster than $1/L$ but slower than $1/L^2$. In the ground states of one-dimensional quantum systems described by short-range Hamiltonians, the spectral gap often scales as $1/L$ at critical points~\cite{cardy1984conformal,cardy1986operator,di1997conformal}. Faster gap closing $O(L^{-k})$ with $k\geq2$ can also occur in particular circumstances, for example, when the Lifshitz transition occurs~\cite{sirker2011J1J2,balents2016quantum,chepiga2023critical,wang2023quantum} or the frustration-free condition is satisfied~\cite{koma1997spectral,gosset2016local,masaoka2024quadratic,lemm2025critical,masaoka2025rigorous,masaoka2026frustration}. The behavior of $\Delta_K$ differs from these conventional scalings commonly encountered in one-dimensional short-range quantum systems. We therefore focus on the gapless phase in the following analysis.

\begin{figure}[tbp]
\begin{center}
\includegraphics[width=6cm]{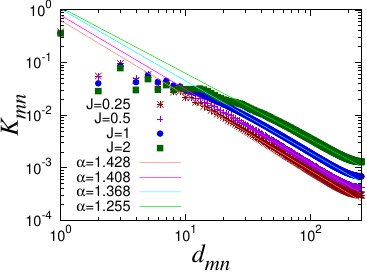}
\caption{The root-mean-square hopping magnitudes of the effective Hamiltonians for various distances, $K_{mn}$ in Eq.~(\ref{eq:intensity-average}), with $\mu_e=\mu_o=0.5$. The time average over $N_s=10^3$ steps is taken after Eq.~\eqref{eq:convergence-condition} is satisfied with $w=3\times10^{-3}$ for all $\nu$. For each $d_{mn}$, we further take the average over $m$ with $n=m+d_{mn}$. Green squares, blue circles, purple crosses, and red asterisks correspond to $J=2$, $1$, $0.5$, and $J=0.25$, respectively. Solid lines depict the algebraic decay $\beta/d_{mn}^\alpha$. The parameters obtained from the least squares method are $(\alpha,\beta)=(1.255,1.113)$ (green), $(1.368,1.040)$ (blue), $(1.408,0.7889)$ (purple), and $(1.428,0.6296)$ (red). The system size is $L=256$, and the fitting is carried out in the range $20 \leq d_{mn} \leq 190$. The deviation from the algebraic decay for large $d_{mn}$ is attributed to finite-size effects. } 
\label{fig:Hamiltonian-magnitude_oshima-model}
\end{center}
\end{figure}

We find that the effective Hamiltonians have algebraically decaying long-range hoppings. Figure~\ref{fig:Hamiltonian-magnitude_oshima-model} shows 
\begin{align}
K_{mn}=\sqrt{\overline{\left|\mathcal{K}_{mn}(\bm{\omega}_t)\right|^2}}
\label{eq:intensity-average}
\end{align}
as a function of the distance between sites $m$ and $n$, $d_{mn}=\min(|n-m|,2L-|n-m|)$. The random hopping magnitudes exhibit algebraic decay with respect to the distance,
\begin{align}
    K_{mn} \propto d_{mn}^{-\alpha}.
    \label{eq:algebraic-decay}
\end{align}
The decay exponent $\alpha$ becomes smaller as the strength of the unitary dynamics increases, as can be seen by comparing the results for $J=0.25$, $0.5$, $1$, and $2$.

\begin{figure}[tbp]
\begin{center}
\includegraphics[width=\columnwidth]{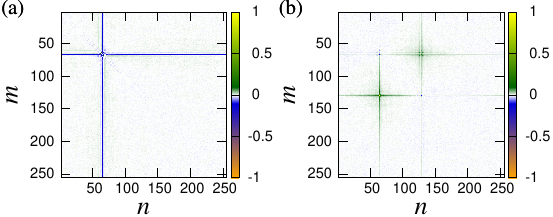}
\caption{The Pearson correlation coefficients between the hopping magnitudes of effective Hamiltonians, $C_{k \ell, mn}^K$ in Eq.~\eqref{eq:correlation_oshima-model}, with $L=128$. The bond specified by $k$ and $\ell$ is fixed as $k=L/2,\ell=L/2+1$ in (a) and  $k=L/2,\ell=L$ in (b). The time average is taken over $N_s=10^5$ steps after Eq.~\eqref{eq:convergence-condition} is satisfied with $w=10^{-3}$ for all $\nu$. The parameters are $J=1$ and $\mu_e=\mu_o=0.5$.} 
\label{fig:Hamiltonian-correlation_oshima-model}
\end{center}
\end{figure}

We also find that the random hoppings are not independently distributed, and their magnitudes exhibit nontrivial correlations. To see this, we compute the Pearson correlation coefficient between the hopping magnitudes $\left|\mathcal{K}_{k\ell}(\bm{\omega}_t)\right|$ and $\left|\mathcal{K}_{mn}(\bm{\omega}_t)\right|$,
\begin{align}
C_{k \ell, mn}^K=\frac{\overline{
\left|\mathcal{K}_{k\ell}(\bm{\omega}_t)\right|\left|\mathcal{K}_{mn}(\bm{\omega}_t)\right|}-\overline{
\left|\mathcal{K}_{k\ell}(\bm{\omega}_t)\right|}\,\overline{\left|\mathcal{K}_{mn}(\bm{\omega}_t)\right|}}{\delta\mathcal{K}_{k\ell}(\bm{\omega}_t)\delta\mathcal{K}_{mn}(\bm{\omega}_t)},
\label{eq:correlation_oshima-model}
\end{align}
where $\delta\mathcal{K}_{k\ell}(\bm{\omega}_t)=\sqrt{\overline{
\left|\mathcal{K}_{k\ell}(\bm{\omega}_t)\right|^2}-\overline{\left|\mathcal{K}_{k\ell}(\bm{\omega}_t)\right|}^2}$. Figure~\ref{fig:Hamiltonian-correlation_oshima-model}(a) shows the correlation $C_{k \ell, mn}^K$ between the nearest-neighbor hopping at $k=L/2,\ell=L/2+1$ and other hoppings. For $m=k$ or $n=\ell$, $\left|\mathcal{K}_{L/2,L/2+1}(\bm{\omega}_t)\right|$ is negatively correlated with $\left|\mathcal{K}_{mn}(\bm{\omega}_t)\right|$. In other words, when the nearest-neighbor hopping magnitude $\left|\mathcal{K}_{L/2,L/2+1}(\bm{\omega}_t)\right|$ is small (large), the other hopping magnitudes sharing one of its endpoints tend to be large (small). By contrast, a different correlation structure emerges for long-range hoppings. Figure~\ref{fig:Hamiltonian-correlation_oshima-model}(b) shows $C_{k\ell,mn}^K$ for a long-range hopping between $k=L/2$ and $\ell=L$. In this case, $\left|\mathcal{K}_{L/2,L}(\bm{\omega}_t)\right|$ and $\left|\mathcal{K}_{mn}(\bm{\omega}_t)\right|$ are positively correlated with endpoints near $m=L/2,n=L$. Thus, when a large (small) long-range hopping magnitude appears in a given realization of $\mathcal{K}(\bm{\omega}_t)$, other nearby long-range hopping magnitudes also tend to be large (small) in the same realization. We note that correlations among hoppings themselves are much weaker than those among hopping magnitudes, as discussed in Appendix~\ref{sec:correlations_K_appendix}.

\subsection{Entanglement scaling}
\label{subsec:entanglement-scaling}
\begin{figure}[tbp]
\begin{center}
\includegraphics[width=6cm]{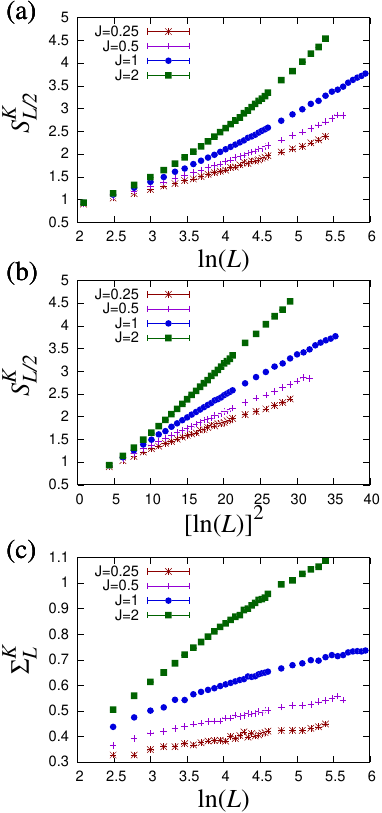}
\caption{The half-chain entanglement entropy of ground states of $\hat{\mathcal{K}}(\bm{\omega}_t)$ as a function of (a) $\ln(L)$ and (b) $[\ln(L)]^2$. (c) $\Sigma_L^K=\left(S_{L/2}^K-S_{L_0/2}^K\right)/\ln\left(L/L_0\right)$ with $L_0=8$ as a function of $\ln(L)$, whose non-zero slope supports a scaling faster than $\ln(L)$. The parameters are $\mu_e=\mu_o=0.5$ and $b=1$. Green squares, blue circles, purple crosses, and red asterisks correspond to $J=2$, $1$, $0.5$, and $0.25$, respectively. The time average is taken over $N_s=10^3$ steps in the long-time regime where $t>4000 \gg L$ is satisfied and thus $\ket{\psi(\bm{\omega}_t)}$ can be regarded as the ground state of $\hat{\mathcal{K}}(\bm{\omega}_t)$.} 
\label{fig:entanglement-entropy_oshima-model}
\end{center}
\end{figure}

We explore the entanglement entropy of quantum trajectories $\ket{\psi(\bm{\omega}_t)}$ generated by the monitored dynamics. To this end, we consider a subsystem $A$ consisting of $2|A|$ consecutive Majorana fermions starting from the site $b$, i.e., $\{\hat{\gamma}_b,\hat{\gamma}_{b+1},\ldots,\hat{\gamma}_{b+2|A|-1}\}$, where the indices are taken modulo $2L$. The Majorana covariance matrix of the subsystem $A$ is given by
\begin{align}
    \Gamma_A(\bm{\omega}_t)=\{\Gamma_{\ell m}(\bm{\omega}_t)\}_{\ell,m \in \{b,b+1,\cdots,b+2|A|-1\}},
    \label{eq:majorana-covariance-matrix_A}
\end{align}
where
\begin{align}
    \Gamma_{\ell m}(\bm{\omega}_t)=\frac{i}{2}\bra{\psi(\bm{\omega}_t)}[\hat{\gamma}_\ell,\hat{\gamma}_m]\ket{\psi(\bm{\omega}_t)}.
\end{align}
The entanglement entropy of the subsystem $A$ can then be obtained from the covariance matrix as
\begin{align}
    S_A(\bm{\omega}_t)=-\frac{1}{2}\mathrm{tr}\left(\frac{\mathbb{I}+i\Gamma_A(\bm{\omega}_t)}{2}\ln\left[\frac{\mathbb{I}+i\Gamma_A(\bm{\omega}_t)}{2}\right]\right),
    \label{eq:entanglement-entropy}
\end{align}
where $\mathbb{I}$ is the $2|A|\times2|A|$ identity matrix.
Figures~\ref{fig:entanglement-entropy_oshima-model}(a) and (b) show $S_{L/2}^K$ as functions of $\ln(L)$ and $[\ln(L)]^2$, respectively, where $S_A^K=\overline{S_A(\bm{\omega}_t)}$ is the time averaged entanglement entropy.
For various parameter sets, which lie within the gapless phase~\cite{oshima2025topology}, we observe that the scaling 
\begin{align}
S_{L/2}^K\propto[\ln(L)]^2
\label{eq:entanglement-scaling}
\end{align}
works better than $S_{L/2}^K\propto\ln(L)$. Figure~\ref{fig:entanglement-entropy_oshima-model}(c) provides further evidence for this scaling. We find that $\Sigma_L^K=\left(S_{L/2}^K-S_{L_0/2}^K\right)/\ln\left(L/L_0\right)$ exhibits a nonzero slope as a function of $\ln(L)$ in agreement with the $[\ln(L)]^2$ scaling in Eq.~\eqref{eq:entanglement-scaling}. 
This scaling is consistent with the field-theoretic prediction for monitored systems with the same parity conservation as that of our model~\cite{fava2023nonlinear}.

This scaling behavior of the entanglement entropy cannot be explained by the corresponding random hopping model with uncorrelated disorder. As discussed in Ref.~\cite{mirlin1996transition}, a random power-law banded model with hopping magnitudes algebraically decaying as $|h_{mn}| \sim 1/d_{mn}^\alpha$ exhibits an Anderson localization transition: it belongs to a delocalized phase for $\alpha<1$ whereas to a localized phase for $\alpha>1$. Additional particle-hole symmetry, as discussed in Appendix~\ref{sec:symmetry}, renders the latter phase delocalized at the band center by singularities in the density of states~\cite{zhou2003one-dimensional}. This delocalized phase for $\alpha>1$ caused by the particle-hole symmetry, akin to random singlet phases in disordered spin chains~\cite{refael2004entanglement}, shows the logarithmic entanglement scaling $S_{L/2} \sim \ln(L)$ whose coefficient varies with $\alpha$~\cite{juhasz2022testing}. For the parameter regime considered here, the effective Hamiltonian for our monitored Majorana chains exhibits $1.2\leq\alpha\leq1.5$; it should belong to the delocalized phase with logarithmic entanglement scaling if the correlations in hopping magnitudes are neglected. Thus, the correlations seem crucial for explaining the observed $[\ln(L)]^2$ scaling in the entanglement entropy.

\section{Long-range correlated models}
\label{sec:long-range-correlated_Hamiltonians}
To understand the relationship among the power-law hoppings in Eq.~\eqref{eq:algebraic-decay}, the non-Gaussian correlations of hopping magnitudes in Fig.~\ref{fig:Hamiltonian-correlation_oshima-model}, and the entanglement scaling in Eq.~\eqref{eq:entanglement-scaling}, we construct a one-dimensional Majorana Hamiltonian,
\begin{align}
    \hat{\mathcal{H}}(\bm{h})=-\frac{i}{4}\vec{\gamma}^\mathsf{T}\mathcal{H}(\bm{h})\vec{\gamma},
\end{align}
where $\mathcal{H}(\bm{h})=\mathcal{H}^*(\bm{h})=-\mathcal{H}^\mathsf{T}(\bm{h})$ is a real antisymmetric matrix. Here, $\bm{h}$ specifies a set of random variables, which will be defined in Sec.~\ref{subsec:model}. Below, we denote by $\overline{X(\bm{h})}$ the average of a quantity $X(\bm{h})$ over these random variables.

The model is designed to reproduce the following properties of the effective Hamiltonians $\mathcal{K}(\bm{\omega}_t)$ found in Sec.~\ref{sec:monitored-majorana}. 
First, the hoppings have random signs with vanishing averages, $\overline{\mathcal{H}_{\ell m}(\bm{h})}=0$, while their root-mean-square magnitudes decay algebraically with distance,
\begin{align}
    \sqrt{\overline{\left|\mathcal{H}_{\ell m}(\bm{h})\right|^2}} \propto d_{\ell m}^{-\chi},
\end{align}
with an exponent $\chi>0$, as in Eq.~\eqref{eq:algebraic-decay}. 
Second, the magnitudes of two hoppings sharing a single site are correlated, as in Fig.~\ref{fig:Hamiltonian-correlation_oshima-model}, where the strength of the correlations is controlled by a small parameter $\eta\geq0$. 
Since $|\mathcal{H}_{\ell m}(\bm{h})|=|\mathcal{H}_{m\ell}(\bm{h})|$, we can take the shared site to be $\ell$ without loss of generality. 
The magnitude of a nearest-neighbor hopping is negatively correlated with that of a longer-range hopping sharing the same site,
\begin{align} \label{eq:property_i}
    \overline{\left|\mathcal{H}_{\ell m}(\bm{h})\right|\left|\mathcal{H}_{\ell,\ell\pm1}(\bm{h})\right|}-\overline{\left|\mathcal{H}_{\ell m}(\bm{h})\right|}\,\overline{\left|\mathcal{H}_{\ell,\ell\pm1}(\bm{h})\right|}\propto -\eta,
\end{align}
for $d_{\ell m}>1$.
By contrast, the magnitudes of two longer-range hoppings sharing the same site are positively correlated, with a strength decaying with the distance between the two unshared sites,
\begin{align} \label{eq:property_ii}
    \overline{\left|\mathcal{H}_{\ell m}(\bm{h})\right|\left|\mathcal{H}_{\ell n}(\bm{h})\right|}-\overline{\left|\mathcal{H}_{\ell m}(\bm{h})\right|}\,\overline{\left|\mathcal{H}_{\ell n}(\bm{h})\right|}\propto +\frac{\eta}{d_{mn}^\kappa},
\end{align}
for $m,n\neq\ell,\ell\pm1$ and $m\neq n,n\pm1$.
Here, we assume that the correlations decay algebraically with an exponent $\kappa>1/2$. 
Equations~\eqref{eq:property_i} and \eqref{eq:property_ii} hold to leading order in $\eta$, as we will discuss below in detail. 
For $\eta=0$, the model reduces to an uncorrelated random power-law hopping model \cite{mirlin1996transition}, which serves as a reference for identifying the role of the correlations.

\subsection{Construction of the model}
\label{subsec:model}
We now describe the explicit construction of the model. We first introduce a real antisymmetric random matrix $h$, which gives the collection of random variables, $\bm{h}=\{h_{\ell m}\}$.
For each off-diagonal pair, we draw
\begin{align}
    h_{\ell m}\sim\mathcal{N}(0,\sigma^2),
    \qquad
    h_{m\ell}=-h_{\ell m},
\end{align}
and set $h_{\ell\ell}=0$.
We define
\begin{align}
    f_{\ell m}=|h_{\ell m}|^2-\sigma^2,
\end{align}
which measures the deviation of the squared hopping magnitude from its mean.
Hereafter, we set $\sigma=1$ and assume the periodic boundary condition.
For hoppings with the distance $d_{\ell m}=\min(|\ell-m|,2L-|\ell-m|)>1$, we define
\begin{align}
    \tilde{h}_{\ell m}=&h_{\ell m}\prod_{s=\pm1}\left(1-\eta f_{\ell,\ell+s}\right)^2\prod_{k \neq\ell,\ell\pm1,m}\left(1+\eta\frac{f_{\ell k}}{d_{km}^\kappa}\right)^2\nonumber\\
    &\times\prod_{s=\pm1}\left(1-\eta f_{m+s,m}\right)^2\prod_{k \neq \ell,m,m\pm1}\left(1+\eta\frac{f_{km}}{d_{k\ell}^\kappa}\right)^2.
    \label{eq:procedure_inducing-correlation}
\end{align}
For nearest-neighbor bonds with $d_{\ell m}=1$, we simply take $\tilde{h}_{\ell m}=h_{\ell m}$, while $\tilde{h}_{\ell\ell}=0$.
The small parameter $\eta\geq0$ controls two types of correlations. 
The products $\prod_{s=\pm1}\left(1-\eta f_{\ell,\ell+s}\right)^2$ and $\prod_{s=\pm1}\left(1-\eta f_{m+s,m}\right)^2$ generate negative correlations between nearest-neighbor and long-range hoppings. 
This is because nearest-neighbor hopping whose intensity $|h_{\ell,\ell\pm1}|^2$ is larger (smaller) than its mean $\sigma^2$ suppresses (enhances) the magnitudes of long-range hoppings sharing the same endpoint. 
In contrast, the factors $\prod_{k \neq\ell,\ell\pm1,m}\left(1+\eta\frac{f_{\ell k}}{d_{km}^\kappa}\right)^2$ and $\prod_{k \neq \ell,m,m\pm1}\left(1+\eta\frac{f_{km}}{d_{k\ell}^\kappa}\right)^2$ enhance (suppress) the magnitude of a long-range hopping when the magnitudes of other long-range hoppings sharing one of its endpoints are larger (smaller) than $\sigma^2$, thereby generating positive correlations among long-range hoppings.
The factor $d_{\ell k}^{-\kappa}$ makes this contribution decay with distance and keeps the variance of $\tilde h_{\ell m}$ finite for $\kappa>0.5$. Finally, we introduce algebraically decaying hoppings,
\begin{align}
    \mathcal{H}_{\ell m}(\bm{h})=\frac{\tilde{h}_{\ell m}}
    {N(\bm{h})\,d_{\ell m}^\chi},
    \label{eq:Hamiltonian_artificial-model}
\end{align}
for $\ell \neq m$, with the normalization 
\begin{align}
    N(\bm{h}) = \sqrt{\frac{1}{L}\sum_{\substack{\ell,m\\ \ell\neq m}}\left|\tilde h_{\ell m}/d_{\ell m}^{\chi}\right|^2}.
\end{align}
Here, $\chi>0$ controls the algebraic decay of the hopping magnitudes. The normalization factor $N(\bm{h})$ is introduced to prevent the norm of $\mathcal{H}(\bm{h})$ from becoming excessively large due to the correlations of hopping magnitudes. With this normalization, the constructed Hamiltonians reproduce the property $\sum_{\ell m}\mathcal{K}^2_{\ell m}(\bm{\omega}_t)/L=O(1)$ satisfied by the effective Hamiltonians in the monitored Majorana chain.
We note that the diagonal elements are zero, $\mathcal{H}_{\ell\ell}(\bm{h})=0$. 

As shown in Appendix~\ref{sec:correlations_H_appendix}, for $m,n\neq\ell,\ell\pm1$ and $m \neq n,n\pm1$, the correlations among hopping magnitudes are given by
\begin{align}
    &\overline{\left|\tilde{h}_{\ell m}\right|\left|\tilde{h}_{\ell \ell+1}\right|}-\overline{\left|\tilde{h}_{\ell m}\right|}\,\overline{\left|\tilde{h}_{\ell \ell+1}\right|}\simeq-\frac{2}{\pi}\eta\left(2-3\eta\right),
    \label{eq:correlation_analytical_negative}\\
    &\overline{\left|\tilde{h}_{\ell m}\right|\left|\tilde{h}_{\ell n}\right|}-\overline{\left|\tilde{h}_{\ell m}\right|}\,\overline{\left|\tilde{h}_{\ell n}\right|}\simeq+\frac{8}{\pi}\eta\left(d_{mn}^{-\kappa}+D_\kappa\eta\right)
    \label{eq:correlation_analytical_positive}
\end{align}
to second order in $\eta$. While the first term in the right-hand side of Eq.~\eqref{eq:correlation_analytical_positive} decays as $d_{mn}$ increases, the second term with $D_\kappa>0$ provides a positive contribution that persists at large distances. These expressions show that $\left|\tilde{h}_{\ell m}\right|$ and $\left|\tilde{h}_{\ell \ell+1}\right|$ are negatively correlated, whereas $\left|\tilde{h}_{\ell m}\right|$ and $\left|\tilde{h}_{\ell n}\right|$ are positively correlated. 
We also note that the magnitudes of two hoppings sharing no site are correlated at most of order $\eta^2$. Thus, the correlations between hopping magnitudes sharing a single site are dominant for small $\eta$.
We assume that the normalization factor does not alter the signs of these correlations, i.e., correlations among $\left|\tilde{h}_{\ell m}\right|$ are preserved in those of $\left|\mathcal{H}_{\ell m}(\bm{h})\right|$.

\begin{figure}[tbp]
\begin{center}
\includegraphics[width=\columnwidth]{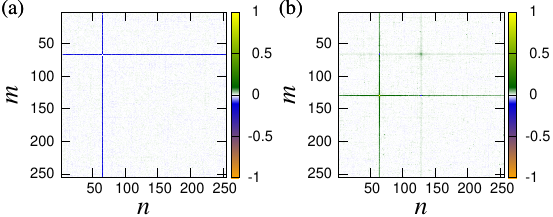}
\caption{The Pearson correlation coefficients of the hopping magnitudes of the constructed Hamiltonians, $C_{k \ell, mn}^H$ in Eq.~\eqref{eq:correlation_artificial-model}, with $L=128$. The bond specified by $k$ and $\ell$ is fixed as $k=L/2,\ell=L/2+1$ in (a) and  $k=L/2,\ell=L$ in (b). The parameters are $\chi=1.3,\eta=0.1$, and $\kappa=0.6$. The average is taken over $10^5$ realizations of $\bm{h}$.} 
\label{fig:Hamiltonian-correlation_artificial-model}
\end{center}
\end{figure}

Figure~\ref{fig:Hamiltonian-correlation_artificial-model} shows the Pearson correlation coefficients
\begin{align}
C_{k \ell, mn}^H=\frac{\overline{
\left|\mathcal{H}_{k\ell}(\bm{h})\right|\left|\mathcal{H}_{mn}(\bm{h})\right|}-\overline{
\left|\mathcal{H}_{k\ell}(\bm{h})\right|}\,\overline{\left|\mathcal{H}_{mn}(\bm{h})\right|}}{\delta\mathcal{H}_{k\ell}(\bm{h})\delta\mathcal{H}_{mn}(\bm{h})},
\label{eq:correlation_artificial-model}
\end{align}
where $\delta\mathcal{H}_{k\ell}(\bm{h})=\sqrt{\overline{
\left|\mathcal{H}_{k\ell}(\bm{h})\right|^2}-\overline{
\left|\mathcal{H}_{k\ell}(\bm{h})\right|}^2}$.  As expected, the constructed model $\mathcal{H}(\bm{h})$ reproduces the essential features of the correlation structure observed in the effective Hamiltonian $\mathcal{K}(\bm{\omega}_t)$; $C_{k \ell, mn}^H$ exhibits negative correlations between nearest-neighbor and long-range hopping magnitudes, as well as positive correlations among long-range hopping magnitudes, similar to Fig.~\ref{fig:Hamiltonian-correlation_oshima-model}.

\subsection{Scaling of gap and entanglement}
\label{subsec:scaling_gap-entanglement}
\begin{figure}[tbp]
\begin{center}
\includegraphics[width=6cm]{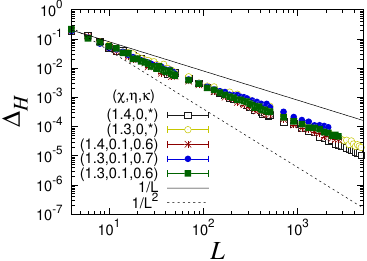}
\caption{The spectral gaps of $\hat{\mathcal{H}}(\bm{h})$ averaged over $10^2$ realizations of $\bm{h}$ for various system sizes. Green squares, blue circles, and red asterisks respectively correspond to $(\chi,\eta,\kappa)=(1.3,0.1,0.6)$, $(1.3,0.1,0.7)$, and $(1.4,0.1,0.6)$. Yellow empty circles and black empty squares respectively correspond to $(\chi,\eta)=(1.3,0)$ and $(1.4,0)$. The black solid and dashed lines are proportional to $1/L$ and $1/L^2$, respectively.} 
\label{fig:gap_artificial-model}
\end{center}
\end{figure}

Figure~\ref{fig:gap_artificial-model} shows the spectral gaps of the normalized Hamiltonian in Eq.~\eqref{eq:Hamiltonian_artificial-model},
\begin{align}
    \Delta_H=2\overline{E_L(\bm{h})},
\end{align}
where $\{\pm iE_1(\bm{h}),\ldots,\pm iE_L(\bm{h})\}$ are single-particle energies of $\mathcal{H}(\bm{h})$ ordered as $E_1(\bm{h}) \geq \ldots \geq E_L(\bm{h}) \geq 0$. As the system size $L$ is increased, the gap decays faster than $1/L$ but slower than $1/L^2$, similar to the behavior of the effective Hamiltonian shown in Fig.~\ref{fig:gap_oshima-model}. We note that the spectral gaps exhibit similar scalings for both $\eta\neq0$ and $\eta=0$. 

\begin{figure}[tbp]
\begin{center}
\includegraphics[width=6cm]{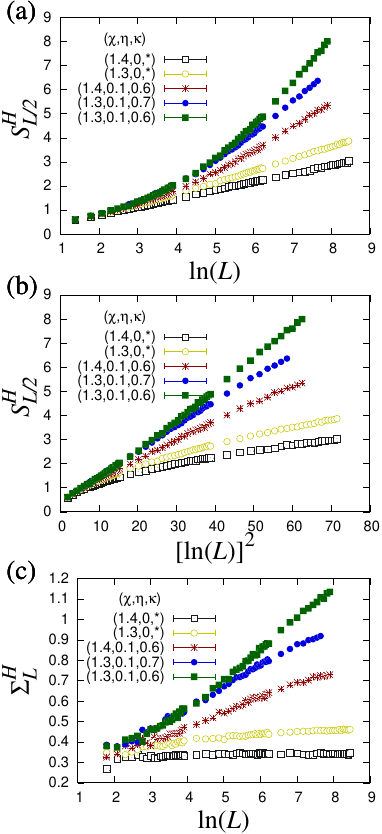}
\caption{The half-chain entanglement entropy of ground states of $\hat{\mathcal{H}}(\bm{h})$ as a function of (a) $\ln(L)$ and (b) $[\ln(L)]^2$. (c) $\Sigma_L^H=\left(S_{L/2}^H-S_{L_0/2}^H\right)/\ln\left(L/L_0\right)$ with $L_0=4$ as a function of $\ln(L)$, whose non-zero slope supports a scaling faster than $\ln(L)$. Green squares, blue circles, and red asterisks respectively correspond to $(\chi,\eta,\kappa)=(1.3,0.1,0.6)$, $(1.3,0.1,0.7)$, and $(1.4,0.1,0.6)$. Yellow empty circles and black empty squares respectively correspond to $(\chi,\eta)=(1.3,0)$ and $(1.4,0)$. The average is taken over $10^2$ realizations of $\bm{h}$ and $10$ different positions of $b$.} 
\label{fig:entanglement-entropy_artificial-model}
\end{center}
\end{figure}

We further explore the ground-state entanglement entropy of $\hat{\mathcal{H}}(\bm{h})$. To this end, we use Eqs.~(\ref{eq:majorana-covariance-matrix_A})-(\ref{eq:entanglement-entropy}) where $\ket{\psi(\bm{\omega}_t)}$ is replaced with the ground state of $\hat{\mathcal{H}}(\bm{h})$.  Figure~\ref{fig:entanglement-entropy_artificial-model} shows the half-chain entanglement entropy $S_{L/2}^H=\overline{S_{L/2}(\bm{h})}$. In the absence of correlations among the hoppings, i.e., for $\eta=0$, the ground-state entanglement entropy exhibits logarithmic scaling as the system size increases,
\begin{align}
    S_{L/2}^H\propto\ln(L).
\end{align}
By contrast, when correlations between different hopping magnitudes are introduced by taking nonzero $\eta$, we observe that the entanglement is enhanced and the scaling seems to obey 
\begin{align}
    S_{L/2}^H\propto[\ln(L)]^2.
\end{align}
Thus, the introduction of correlations between hopping magnitudes can change the entanglement scaling from $\ln(L)$ to $[\ln(L)]^2$. This result suggests that correlations among long-range hopping magnitudes play a crucial role in generating the unconventional entanglement scaling $[\ln(L)]^2$ observed in the monitored Majorana dynamics. 
However, precise matching between the effective Hamiltonian for monitored Majorana circuits and the model parameters or model itself used in this analysis is left for future work.

\section{Conclusion}
\label{sec:conclusion}
We have investigated the structure of effective Hamiltonians governing monitored Majorana dynamics. We have shown that the effective Hamiltonians $\mathcal{K}(\bm{\omega}_t)$ exhibit random power-law hoppings with nontrivial correlations: nearest-neighbor and long-range hopping magnitudes are negatively correlated, whereas different long-range hopping magnitudes are positively correlated. We have also found that, in the gapless phase, the spectral gap closes faster than $1/L$ but slower than $1/L^2$, while the ground-state entanglement entropy scales as $[\ln(L)]^2$. These behaviors are unusual compared with those commonly encountered in the ground states of short-range interacting systems, e.g., at conformal or Lifshitz critical points.

To elucidate the role of the non-Gaussian correlations among hoppings, we have constructed random Hamiltonians $\mathcal{H}(\bm{h})$ with power-law hoppings that capture the essential features of the effective Hamiltonians $\mathcal{K}(\bm{\omega}_t)$, namely, negative correlations between nearest-neighbor and long-range hopping magnitudes and positive correlations among long-range hopping magnitudes. We have found that the spectral gap of the constructed models also closes faster than $1/L$ but slower than $1/L^2$ as the system size increases. More importantly, the ground-state entanglement entropy is consistent with $[\ln(L)]^2$ scaling in the presence of these hopping correlations, whereas it exhibits $\ln(L)$ scaling when the correlations are absent. These results suggest that correlations among long-range hopping magnitudes play a key role in generating the unconventional $[\ln(L)]^2$ entanglement scaling observed in the monitored Majorana dynamics. 

\begin{acknowledgments}
We thank Zongping Gong for valuable discussions.
K. M. and R.H. are supported by JST ERATO Grant Number JPMJER2302, Japan. 
K.M. is supported by JSPS KAKENHI Grant No.~JP23K13037.
R.H. is supported by JSPS KAKENHI Grant No.~JP24K16982.
Y.F. is supported by JSPS KAKENHI Grant No.~JP24K06897. 
We used generative AI tools (ChatGPT Pro 5.6 and Claude Opus 5.5) to improve the presentation of this manuscript. Interactions with these tools helped inspire the idea of the proofs in Appendices. All results were independently verified by the authors, who take full responsibility for the content of this paper.
\end{acknowledgments}
\appendix

\section{Symmetries of the effective Hamiltonians}
\label{sec:symmetry}
We investigate symmetries of the effective Hamiltonians. For this purpose, we review the construction of effective Hamiltonians in the complex-fermion representation. After that, we examine how the symmetries of the Kraus and unitary time-evolution operators are inherited by the effective Hamiltonians. 

The fermionic annihilation operators are related to the Majorana operators by
\begin{align}
    \hat{c}_j=\frac{\hat{\gamma}_{2j-1}+i\hat{\gamma}_{2j}}{2},
\end{align}
where $\{\hat{c}_i,\hat{c}_j^\dagger\}=\delta_{ij}$ and $\{\hat{c}_i,\hat{c}_j\}=0$ are satisfied. Thus, a Majorana chain with $2L$ modes is mapped onto a complex-fermion chain with $L$ sites. In this representation, the effective Hamiltonian can be written as
\begin{align}
    \hat{\mathcal{K}}(\bm{\omega}_t)=\frac{1}{2}\vec{c}^\dagger\tilde{\mathcal{K}}(\bm{\omega}_t)\vec{c}+c\hat{\mathbb{I}},
\end{align}
where $\vec{c}=(\hat{c}_1,\ldots,\hat{c}_L,\hat{c}_1^\dagger,\ldots,\hat{c}_L^\dagger)^\mathsf{T}$. To derive $\tilde{\mathcal{K}}(\bm{\omega}_t)$, we first express the elementary operators constituting $\hat{\mathcal{M}}(\bm{\omega}_t)$ in the complex-fermion representation. The unitary evolution operator in Eq.~(\ref{eq:unitary-operator}) can be written as
\begin{align}
    \hat{U}&=\exp\left(-i\vec{c}^\dagger S \vec{c}\right)\nonumber\\
    &=\exp\left(iJ \sum_{j=1}^{L} \hat{c}_j^\dagger\hat{c}_{j+1}+\hat{c}_j^\dagger\hat{c}_{j+1}^\dagger-\hat{c}_j^\dagger\hat{c}_j+h.c.\right),
\end{align}
where $S$ is a Hermitian matrix and constant terms are omitted. Under this unitary evolution, the creation and annihilation operators are transformed as
\begin{align}
    \hat{U}^\dagger\vec{c}\hat{U}=\exp(-2iS)\vec{c}.
\end{align}
The Kraus operators for Majorana-parity measurements on odd bonds, given in Eq.~(\ref{eq:Kraus-operator_odd}), take the form
\begin{align}
\hat{M}^o_j(\omega)&=\frac{\exp\left[\vec{c}^\dagger T_j^o(\omega)\vec{c}\right]}{\sqrt{2\cosh (2\theta_o)}},
\label{eq:Kraus-operator_odd_appendix}
\end{align}
in the complex-fermion representation. Here, $T_j^o(\omega)$ is given by
\begin{align}
    [T_j^o(\omega)]_{k\ell}=\omega\theta_o(-\delta_{j,k}\delta_{j,\ell}+\delta_{j+L,k}\delta_{j+L,\ell}).
    \label{eq:T_odd}
\end{align}
The non-zero eigenvalues of $T_j^o(\omega)$ become $\pm\theta_o$. Similarly, the Kraus operators for Majorana-parity measurements on even bonds, given in Eq.~(\ref{eq:Kraus-operator_even}), can be written as
\begin{align}
\hat{M}^e_j(\omega)&=\frac{\exp\left[\vec{c}^\dagger T_j^e(\omega)\vec{c}\right]}{\sqrt{2\cosh (2\theta_e)}},
\label{eq:Kraus-operator_even_appendix}
\end{align}
where
\begin{align}
    [T_j^e(\omega)]_{k\ell}=\frac{\omega\theta_e}{2}[&\delta_{j,k}\delta_{j+1,\ell}+\delta_{j,k}\delta_{j+L+1,\ell}-\delta_{j+1,k}\delta_{j+L,\ell}\nonumber\\
    &-\delta_{j+L,k}\delta_{j+L+1,\ell}+(k\leftrightarrow\ell)].
    \label{eq:T_even}
\end{align}
The non-zero eigenvalues of $T_j^e(\omega)$ become $\pm\theta_e$. Under the action of the Kraus operators, the fermionic operators transform according to
\begin{align}
[\hat{M}_j^{o/e}(\omega)]^\dagger\vec{c}\left\{[\hat{M}^{o/e}_j(\omega)]^\dagger\right\}^{-1}=\exp[-2T_j^{o/e}(\omega)]\vec{c}.
\end{align}
The single-particle effective Hamiltonian in the complex-fermion representation is therefore given by
\begin{align}
    \tilde{\mathcal{K}}(\bm{\omega}_t)=\frac{1}{2t}\ln[\tilde{\mathcal{M}}(\bm{\omega}_t)\tilde{\mathcal{M}}^\dagger(\bm{\omega}_t)],
    \label{eq:effective-Hamiltonian_complex-fermion}
\end{align}
where
\begin{align}
    \tilde{\mathcal{M}}(\bm{\omega}_t)=\prod_{s=1}^t\left[\prod_je^{-2T_j^e(\omega_{2j,s})}\right]\left[\prod_je^{-2T_j^o(\omega_{2j-1,s})}\right]e^{-2iS}.
    \label{eq:time-evolution-operator_complex-fermion}
\end{align}
Here, the constant term is omitted. The eigenenergies of $\tilde{\mathcal{K}}(\bm{\omega}_t)$ occur in pairs, $\{\pm\varepsilon_1(\bm{\omega}_t),\ldots,\pm\varepsilon_L(\bm{\omega}_t)\}$, where $\varepsilon_i(\bm{\omega}_t)\geq\varepsilon_{i+1}(\bm{\omega}_t)\geq0$ is satisfied.

We first consider the measurement-only dynamics with $J=0$. The matrices $T_j^{o}(\omega)$ and $T_j^{e}(\omega)$, defined in Eqs.~\eqref{eq:T_odd} and \eqref{eq:T_even}, respectively, are real and satisfy
\begin{align}
    [T_j^{o/e}(\omega)]^*=T_j^{o/e}(\omega).
    \label{eq:time-reversal-symmetry_T}
\end{align}
Consequently, the single-particle time-evolution operator defined in Eq.~\eqref{eq:time-evolution-operator_complex-fermion} satisfies
\begin{align}
    \tilde{\mathcal{M}}^*(\bm{\omega}_t)=\tilde{\mathcal{M}}(\bm{\omega}_t).
    \label{eq:time-reversal-symmetry_M}
\end{align}
Equation~\eqref{eq:time-reversal-symmetry_M} implies time-reversal symmetry of the effective Hamiltonian,
\begin{align}
    \tilde{\mathcal{K}}^*(\bm{\omega}_t)=\tilde{\mathcal{K}}(\bm{\omega}_t).
    \label{eq:time-reversal-symmetry_K}
\end{align}
The matrices $T_j^{o}(\omega)$ and $T_j^{e}(\omega)$ also satisfy
\begin{align}
    \tau_xT_j^{o/e}(\omega)\tau_x=-T_j^{o/e}(\omega),
    \label{eq:chiral-symmetry_T}
\end{align}
where $\tau_x=\mathbb{I}\otimes\sigma_x$ with $\mathbb{I}$ being the $L \times L$ identity matrix. This relation connects the time-evolution operator to its inverse Hermitian conjugate as
\begin{align}
    \tau_x\tilde{\mathcal{M}}(\bm{\omega}_t)\tau_x=\left[\tilde{\mathcal{M}}^{-1}(\bm{\omega}_t)\right]^\dagger.
    \label{eq:chiral-symmetry_M}
\end{align}
It follows that the effective Hamiltonian has chiral symmetry,
\begin{align}
    \tau_x\tilde{\mathcal{K}}(\bm{\omega}_t)\tau_x=-\tilde{\mathcal{K}}(\bm{\omega}_t)
    \label{eq:chiral-symmetry_K}.
\end{align}
Combining Eqs.~\eqref{eq:time-reversal-symmetry_M} and \eqref{eq:chiral-symmetry_M}, we obtain
\begin{align}
    \tau_x\tilde{\mathcal{M}}^*(\bm{\omega}_t)\tau_x=\left[\tilde{\mathcal{M}}^{-1}(\bm{\omega}_t)\right]^\dagger,
    \label{eq:particle-hole-symmetry_M}
\end{align}
which gives particle-hole symmetry of the effective Hamiltonian,
\begin{align}
    \tau_x\tilde{\mathcal{K}}^*(\bm{\omega}_t)\tau_x=-\tilde{\mathcal{K}}(\bm{\omega}_t).
    \label{eq:particle-hole-symmetry_K}
\end{align}
Thus, the effective Hamiltonians for the measurement-only dynamics belong to symmetry class BDI.

We next consider the dynamics with $J\neq0$. The matrix $S$ satisfies
\begin{align}
    S^*=S,\ \ 
    \tau_xS\tau_x=-S,\ \ 
    \tau_xS^*\tau_x=-S.
    \label{eq:symmetries_S}
\end{align}
The first and second relations imply
\begin{align}
    \left(e^{-2iS}\right)^*=e^{+2iS},\ \ \tau_x\left(e^{-2iS}\right)\tau_x=e^{+2iS},
\end{align}
respectively. The unitary factor $e^{-2iS}$ therefore breaks the time-reversal and chiral symmetry conditions in Eqs.~\eqref{eq:time-reversal-symmetry_M} and \eqref{eq:chiral-symmetry_M}. By contrast, the third relation in Eq.~\eqref{eq:symmetries_S} gives
\begin{align}
    \tau_x\left(e^{-2iS}\right)^*\tau_x=e^{-2iS},
\end{align}
so that the particle-hole symmetry in Eq.~\eqref{eq:particle-hole-symmetry_K} remains intact. Therefore, for nonzero $J$, the effective Hamiltonians belong to symmetry class D. We note that a similar symmetry analysis has been carried out in Ref.~\cite{xiao2026symmetry}.

\section{Average of $\mathcal{K}(\bm{\omega}_t)$}
\label{sec:Hamiltonian-average}
We show that the ensemble average and time average of the single-particle effective Hamiltonian $\mathcal{K}(\bm{\omega}_t)$ vanish. We first focus on the ensemble average. We consider the transformation of the time-evolution operator under the antiunitary operator
\begin{align} \label{eq:transformationV}
\hat{V}=\left(\prod_{j=1}^{L}\hat{\gamma}_{2j}\right)C,
\end{align}
where $L$ is assumed to be even. Here, $C$ is the complex conjugation in the conventional basis where the matrix representations of fermionic creation and annihilation operators are real.
The Majorana operators are invariant under $\hat{V}$,
\begin{align}
\hat{V}\hat{\gamma}_{\ell}\hat{V}^{-1}=\hat{\gamma}_{\ell}, 
\label{eq:transformation_V-gamma}
\end{align}
which follows from the canonical anticommutation relation $\{\hat{\gamma}_\ell,\hat{\gamma}_m\}=2\delta_{\ell m}$ and the representation $\hat{c}_j=\hat{c}_j^*=(\hat{\gamma}_{2j-1}+i\hat{\gamma}_{2j})/2$, where $\hat{c}_j$ is the fermionic annihilation operator.
Applying the transformation in Eq.~\eqref{eq:transformationV} to Eqs.~\eqref{eq:unitary-operator}, \eqref{eq:Kraus-operator_odd}, and \eqref{eq:Kraus-operator_even} yields
\begin{align}
\hat{V}\hat{U}\hat{V}^{-1}=\hat{U},\ \ \hat{V}\hat{M}_j^{o/e}(\omega)\hat{V}^{-1}=\hat{M}_j^{o/e}(-\omega).
\label{eq:transformation_V-UM}
\end{align}
Thus, the $t$-step time-evolution operator in Eq.~\eqref{eq:Kraus-operator_t-step_many-body} transforms as
\begin{align}
\hat{V} \hat{\mathcal{M}}(\bm{\omega}_t) \hat{V}^{-1} = \hat{\mathcal{M}}(-\bm{\omega}_t),
\end{align}
where $-\bm{\omega}_t$ is the measurement record obtained by flipping the sign of every measurement outcome in $\bm{\omega}_t$.
As a result, the many-body effective Hamiltonian defined in Eq.~\eqref{eq:effective-hamiltonian} transforms as
\begin{align}
\hat{V}\hat{\mathcal{K}}(\bm{\omega}_t)\hat{V}^{-1}=\hat{\mathcal{K}}(-\bm{\omega}_t).
\label{eq:transformation_V-K}
\end{align}

The single-particle effective Hamiltonian $\mathcal{K}(\bm{\omega}_t)$ is related to the many-body effective Hamiltonian $\hat{\mathcal{K}}(\bm{\omega}_t)$ via Eq.~\eqref{eq:effham-single-to-many} with the constant shift $c$ explicitly given by
\begin{align}
    c &= -\frac{1}{2^{L+1}t}\mathrm{tr}
    \left\{
    \ln\left[
    \hat{\mathcal{M}}(\bm{\omega}_t)
    \hat{\mathcal{M}}^\dagger(\bm{\omega}_t)
    \right]
    \right\}\nonumber\\
    &=
    \frac{L}{2}
    \left\{
    \ln\left[2\cosh(2\theta_{\mathrm{o}})\right]
    +
    \ln\left[2\cosh(2\theta_{\mathrm{e}})\right]
    \right\}
    \nonumber\\
    &=
    \frac{L}{2}
    \left\{
    \ln\left[\frac{2(1+\mu_{\mathrm{o}}^2)}{1-\mu_{\mathrm{o}}^2}\right] + \ln\left[\frac{2(1+\mu_{\mathrm{e}}^2)}{1-\mu_{\mathrm{e}}^2}\right]
    \right\}.
    \label{eq:constant_shift}
\end{align}
In this expression, only the normalization factors of the Kraus operators in Eqs.~\eqref{eq:Kraus-operator_odd} and
\eqref{eq:Kraus-operator_even} appear, and thus $c$ is independent of the measurement record $\bm{\omega}_t$.
Since $\hat{V}$ is antiunitary and satisfies Eq.~\eqref{eq:transformation_V-gamma}, by virtue of Eq.~\eqref{eq:transformation_V-K}, we find
\begin{align}
    \hat{\mathcal{K}}(-\bm{\omega}_t) = \frac{i}{4}\vec{\gamma}^\mathsf{T}\mathcal{K}(\bm{\omega}_t)\vec{\gamma} + c\hat{\mathbb{I}}.
\end{align}
Comparing this with
\begin{align}
    \hat{\mathcal{K}}(-\bm{\omega}_t) = - \frac{i}{4}\vec{\gamma}^\mathsf{T}\mathcal{K}(-\bm{\omega}_t)\vec{\gamma} + c\hat{\mathbb{I}},
\end{align}
we obtain
\begin{align}
    \mathcal{K}(-\bm{\omega}_t) = -\mathcal{K}(\bm{\omega}_t).
\end{align}

Here, we take the initial state as
\begin{align}
\hat{\rho}_0=\frac{\hat{P}_+}{\mathrm{tr}\left(\hat{P}_+\right)},
\label{eq:initial-state}
\end{align}
where $\hat{P}_+$ is the projection operator onto the even parity sector.
For even $L$, $\hat{V}$ preserves fermion parity, so that
\begin{align}
    \hat{V}\hat{\rho}_0\hat{V}^{-1} = \hat{\rho}_0.
\end{align}
The Born probability,
\begin{align}
    p(\bm{\omega}_t) = \mathrm{tr}\left[\hat{\mathcal{M}}(\bm{\omega}_t)\hat{\rho}_0\hat{\mathcal{M}}^\dagger(\bm{\omega}_t)\right],
\end{align}
then satisfies
\begin{align}
p(\bm{\omega}_t)=\mathrm{tr}\left[\hat{V}\hat{\mathcal{M}}(\bm{\omega}_t)\hat{\rho}_0\hat{\mathcal{M}}^\dagger(\bm{\omega}_t)\hat{V}^{-1}\right]=p(-\bm{\omega}_t).
\label{eq:transformation_V-p}
\end{align}
Therefore, pairing each measurement record with its sign-flipped counterpart gives 
\begin{align}
\sum_{\bm{\omega}_t}p(\bm{\omega}_t)\mathcal{K}(\bm{\omega}_t)&=\frac{1}{2}\sum_{\bm{\omega}_t}p(\bm{\omega}_t)\left[\mathcal{K}(\bm{\omega}_t)+\mathcal{K}(-\bm{\omega}_t)\right]
    =0.
    \label{eq:mean-zero_ensemble-average}
\end{align}
Thus, the single-particle effective Hamiltonians have a vanishing ensemble average for the probability measure generated from $\hat{\rho}_0$. 

As shown in Appendix~\ref{sec:ergodicity}, the averaged CPTP dynamics corresponding to our monitored Majorana circuit has a unique positive-definite steady state in each parity sector. This ensures that the probability measure of the measurement record generated from $\hat{\rho}_0$ becomes invariant and ergodic~\cite{benoist2019invariant,hamazaki2026an}. Consequently, the long-time statistics are governed by a unique invariant measure of the measurement record for any initial state in the same parity sector, including the vacuum state considered in the main text. It follows that the ensemble average of $\mathcal{K}(\bm{\omega}_t)$ vanishes in the long-time limit for any initial state within the even-parity sector.

As discussed in Refs.~\cite{benoist2019invariant,hamazaki2026an}, the existence of a unique steady state, together with the purification condition~\cite{Maasen2006purification}, ensures that the long-time average along a single trajectory coincides with the ensemble average over different trajectories for quantities nonlinear in the density matrix. For our monitored Majorana circuit, the purification condition is satisfied as discussed in Sec.~7.3.2 of Ref.~\cite{hamazaki2026an}. This allows us to compute the time average of the entanglement entropy only from a single trajectory as done in the main text. However, to the best of our knowledge, such correspondence between the long-time and ensemble average has not been known for the elements of the effective Hamiltonian $\mathcal{K}(\bm{\omega}_t)$ and their correlations. We therefore assume that this correspondence holds for our monitored circuit. Indeed, we have numerically confirmed that the long-time average of $\mathcal{K}(\bm{\omega}_t)$ vanishes even for a single trajectory generated from the vacuum state, while the data are not shown, as expected from the above discussion for the ensemble average of $\mathcal{K}(\bm{\omega}_t)$.

\section{Uniqueness of the stationary state in each parity sector}
\label{sec:ergodicity}
We prove that Eq.~\eqref{eq:initial-state} is the unique stationary state of the averaged CPTP dynamics in the even-parity sector for $\mu_e,\mu_o\neq0$. We can easily extend the discussions below to the case of the odd-parity sector where the unique stationary state is $\left(\hat{\mathbb{I}}-\hat{P}_+\right)/\mathrm{tr}\left(\hat{\mathbb{I}}-\hat{P}_+\right)$. Since the CPTP dynamics is unital and preserves fermion parity, $\hat{\rho}_0$ is a stationary state: $\sum_{\bm{\omega}_t}\hat{\mathcal{M}}(\bm{\omega}_t)\hat{\rho}_0\hat{\mathcal{M}}^\dagger(\bm{\omega}_t)=\hat{\rho}_0$. The necessary and sufficient condition for the unique positive-definite stationary state in the even-parity sector is 
\begin{align}
    \mathsf{A}_+\left\{\hat{\mathbb{I}},\hat{M}(\omega_t)\right\}=\mathbb{B}[\mathscr{H}_+],
    \label{eq:algebra_M-H}
\end{align}
where $\mathbb{B}[\mathscr{H}_+]$ is the set of bounded operators in the even-parity sector $\mathscr{H}_+$ and $\mathsf{A}_+\{\hat{\mathbb{I}},\hat{M}(\omega_t)\}$ is the algebra generated by the identity and one-step time-evolution operators $\hat{M}(\omega_t)$ restricted to $\mathscr{H}_+$~\cite{hamazaki2026an}. Thus, if Eq.~\eqref{eq:algebra_M-H} is satisfied, the positive-definite stationary state $\hat{\rho}_0$ is unique.

We prove Eq.~\eqref{eq:algebra_M-H} by explicitly constructing the complete orthogonal basis for $\mathbb{B}[\mathscr{H}_+]$. 
It suffices to consider the $J=0$ case, as it can be readily extended to the $J \neq 0$ case. 
Since 
\begin{align}
    \sum_{\omega_t}\omega_{\ell,t}\hat{M}(\omega_t)\propto\hat{\gamma}_\ell\hat{\gamma}_{\ell+1}
    \label{eq:M-QU}
\end{align}
is satisfied, the set of products of neighboring Majorana operators, $\{\hat{P}_+\hat{\gamma}_\ell\hat{\gamma}_{\ell+1}\hat{P}_+\}$, is a subset of $\mathsf{A}_+\left\{\hat{\mathbb{I}},\hat{M}(\omega_t)\right\}$. 
By using $(\hat{\gamma}_\ell \hat{\gamma}_{\ell+1}) (\hat{\gamma}_{\ell+1} \hat{\gamma}_{\ell+2})\cdots(\hat{\gamma}_{m-1} \hat{\gamma}_m) = \hat{\gamma}_\ell \hat{\gamma}_m$ and taking their products, we can generate all even-degree monomials of Majorana operators, $\hat{G}_X = \hat{P}_+ \hat{\gamma}_{\ell_1} \hat{\gamma}_{\ell_2} \cdots \hat{\gamma}_{\ell_{2m}} \hat{P}_+$ with $X=\{\ell_1,\cdots\ell_{2m}\}$.
These monomials $\hat{G}_X$ are orthogonal operators in $\mathbb{B}[\mathscr{H}_+]$: for $X'\notin\{ X,\bar{X}\}$, where $\bar{X}$ is the complement of $X$, the Hilbert-Schmidt inner product of $\hat{G}_X$ and $\hat{G}_{X'}$ satisfies 
\begin{align}
    \mathrm{tr}\left(\hat{P}_+\hat{G}^\dagger_{X'}\hat{G}_X\right)=0.
    \label{eq:orthogonal_GX}
\end{align}
In addition, they form the complete basis for $\mathbb{B}[\mathscr{H}_+]$ since the number of independent $\hat{G}_X$ is given by
\begin{align}
    \frac{1}{2}\sum_{m=0}^L\left(\begin{array}{c}2L\\2m\end{array}\right)=2^{2L-2}=[\dim(\mathscr{H}_+)]^2,
    \label{eq:number-of-basis}
\end{align}
where the factor $1/2$ compensates the double counting of $\hat{G}_X \propto \hat{G}_{\bar{X}}$. Equations~\eqref{eq:orthogonal_GX} and \eqref{eq:number-of-basis} imply Eq.~\eqref{eq:algebra_M-H}. For $J\neq0$, the right-hand side of Eq.~\eqref{eq:M-QU} is replaced by $\hat{\gamma}_\ell \hat{\gamma}_{\ell+1} \hat{U}$. However, the factor $\hat{U}$ can be canceled and the discussion reduces to the case $J=0$ above. To see this, we notice that $\hat{U}$ is an element of $\mathsf{A}_+\left\{\hat{\mathbb{I}},\hat{M}(\omega_t)\right\}$ owing to $\sum_{\omega_t}\hat{M}(\omega_t)\propto\hat{U}$. The Cayley-Hamilton theorem ensures that $\hat{U}^{-1}$ can be expressed as a finite-degree polynomial of $\hat{U}$, and thus $\hat{P}_+\hat{U}^{-1}\hat{P}_+$ is also an element of $\mathsf{A}_+\left\{\hat{\mathbb{I}},\hat{M}(\omega_t)\right\}$.  It follows that $\{\hat{\gamma}_\ell\hat{\gamma}_{\ell+1}\}$ is a subset of $\mathsf{A}_+\left\{\hat{\mathbb{I}},\hat{M}(\omega_t)\right\}$, leading to Eq.~\eqref{eq:algebra_M-H}.

\section{Upper bound on the spectral width of effective Hamiltonians $\mathcal{K}(\bm{\omega}_t)$}
\label{sec:spectral-width}
We show that the width of the single-particle Lyapunov spectrum is bounded independently of the system size. To this end, we consider the majorization and log-majorization for arrays of real numbers. Consider two arrays $\bm{u}=(u_1,u_2,\cdots,u_N)$ and $\bm{v}=(v_1,v_2,\cdots,v_N)$ whose components are arranged in nonincreasing order. The array $\bm{u}$ is said to be majorized by $\bm{v}$, denoted as
\begin{align}
    \bm{u} \prec \bm{v},
\end{align}
if
\begin{align}
    \sum_{i=1}^nu_i \leq \sum_{i=1}^nv_i
\end{align}
is satisfied, where the equality is achieved for $n=N$. Equivalently, $\bm{u}\prec\bm{v}$ if and only if there exists a doubly stochastic matrix $W$ such that
\begin{align}
    \bm{u}=W\bm{v}.
\end{align}
Here, a doubly stochastic matrix is defined as a non-negative matrix satisfying 
\begin{align}
\sum_iW_{ij}=\sum_iW_{ji}=1
\end{align} 
for arbitrary $j$. For arrays with non-negative components, the log majorization, denoted as
\begin{align}
    \bm{u}\prec_\mathrm{log}\bm{v},
\end{align} 
is defined as
\begin{align}
    \prod_{i=1}^nu_i \leq \prod_{i=1}^nv_i, 
\end{align}
where the equality is satisfied for $n=N$. We apply the log majorization to the singular values of matrices. Given an $N \times N$ matrix $A$, we consider the array $\bm{s}(A)=[s_1(A),s_2(A),\cdots,s_N(A)]$, where $\{s_i(A)\}$ represent the singular values of $A$ and $s_i(A) \geq s_{i+1}(A)$ is satisfied. The Horn theorem states that the singular values of a product of two matrices satisfy
\begin{align}
    \bm{s}(AB) \prec_\mathrm{log} \bm{s}(A)\bm{s}(B),
    \label{eq:Horn-theorem}
\end{align}
where $\bm{s}(A)\bm{s}(B)=[s_1(A)s_1(B),\cdots,s_N(A)s_N(B)]$~\cite{hiai2024log}. 

Applying Eq.~(\ref{eq:Horn-theorem}) repeatedly to the product in Eq.~(\ref{eq:time-evolution-operator_complex-fermion}), we obtain
\begin{align}
    \bm{m}(\bm{\omega}_t)\prec_\mathrm{log}[e^{+2(\theta_e+\theta_o)t},\ldots,e^{-2(\theta_e+\theta_o)t}],
    \label{eq:log-majorization_M}
\end{align}
where $\bm{m}(\bm{\omega}_t)$ denotes the array of singular values of $\tilde{\mathcal{M}}(\bm{\omega}_t)$. We note that singular values of $\prod_je^{-2T_j^o(\omega_{2j-1,s})}$ and $\prod_je^{-2T_j^e(\omega_{2j,s})}$ can be obtained from eigenvalues of the Hermitian matrices $T_j^{o/e}(\omega)$, since $[T_j^o(\omega),T_k^o(\omega')]=[T_j^e(\omega),T_k^e(\omega')]=0$ is satisfied. Taking the logarithm of Eq.~(\ref{eq:log-majorization_M}) and dividing by $t$, we obtain the corresponding majorization relation for the Lyapunov spectrum,
\begin{align}
    \bm{\varepsilon}\prec\bm{\theta},
\end{align}
where $\bm{\varepsilon}=(+\varepsilon_1,\ldots,+\varepsilon_L,-\varepsilon_L,\ldots,-\varepsilon_1)$ and $\bm{\theta}=[+2(\theta_e+\theta_o),\ldots,-2(\theta_e+\theta_o)]$. It follows from the majorization relation that there exists a doubly stochastic matrix $W$ such that
\begin{align}
\bm{\varepsilon}=W\bm{\theta}.
\end{align}
In particular, the largest Lyapunov exponent $\varepsilon_1$ is a convex combination of the components of $\bm{\theta}$. Therefore, the width of the Lyapunov spectrum satisfies
\begin{align}
    2\varepsilon_1\leq4(\theta_e+\theta_o),
\end{align}
independently of the system size $L$. Thus, both the spectral width and the operator norm of the single-particle effective Hamiltonian $\tilde{\mathcal{K}}(\bm{\omega}_t)$ remain bounded in the thermodynamic limit.

\section{Correlations among hoppings of $\mathcal{K}(\bm{\omega}_t)$}
\label{sec:correlations_K_appendix}
\begin{figure}[tbp]
\begin{center}
\includegraphics[width=\columnwidth]{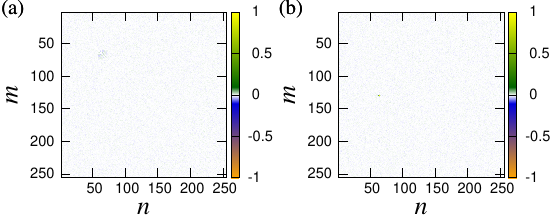}
\caption{The Pearson correlation coefficients of the hoppings, $\tilde{C}_{k \ell, mn}^K$ in Eq.~(\ref{eq:correlation_oshima-model_appendix}), with $L=128$. In (a), $k$ and $\ell$ are fixed as $k=L/2,\ell=L/2+1$. In (b), $k$ and $\ell$ are fixed as $k=L/2,\ell=L$. The parameters are $J=1$ and $\mu_e=\mu_o=0.5$.} 
\label{fig:Hamiltonian-correlation_oshima-model_appendix}
\end{center}
\end{figure}

We explore correlations among hoppings of $\mathcal{K}(\bm{\omega}_t)$, through the Pearson correlation coefficient,
\begin{align}
\tilde{C}_{k \ell, mn}^K=\frac{\overline{
\mathcal{K}_{k\ell}(\bm{\omega}_t)\mathcal{K}_{mn}(\bm{\omega}_t)}-\overline{
\mathcal{K}_{k\ell}(\bm{\omega}_t)}\,\overline{\mathcal{K}_{mn}(\bm{\omega}_t)}}{\delta\mathcal{K}_{k\ell}(\bm{\omega}_t)\delta\mathcal{K}_{mn}(\bm{\omega}_t)}.
\label{eq:correlation_oshima-model_appendix}
\end{align}
Unlike $C_{k\ell,mn}^{K}$ defined in the main text, which measures correlations between the magnitudes of the hoppings, $\tilde{C}_{k\ell,mn}^{K}$ measures correlations between the hoppings themselves. 
Figure \ref{fig:Hamiltonian-correlation_oshima-model_appendix} shows $\tilde{C}_{k \ell, mn}^K$ as functions of $m$ and $n$, where $\mathcal{K}_{k\ell}(\bm{\omega}_t)$ is fixed to either a nearest-neighbor or long-range hopping. Although weak correlations are observed between hoppings located close to each other, the correlations are close to zero for most pairs. Comparing Figs. \ref{fig:Hamiltonian-correlation_oshima-model} and \ref{fig:Hamiltonian-correlation_oshima-model_appendix}, we can see that correlations among $|\mathcal{K}_{k\ell}(\bm{\omega}_t)|$ are much stronger than those among $\mathcal{K}_{k\ell}(\bm{\omega}_t)$, which suggests that the latter is negligible relative to the former. 

\begin{widetext}
\section{Correlations among hopping magnitudes of $\mathcal{H}(\bm{h})$}
\label{sec:correlations_H_appendix}
We investigate the correlations among the hopping magnitudes in the constructed model described by $\mathcal{H}(\bm{h})$. Assuming that the normalization factor does not significantly affect these correlations, we analyze $\overline{\left|\tilde{h}_{\ell m}\right|\left|\tilde{h}_{\ell n}\right|}-\overline{\left|\tilde{h}_{\ell m}\right|}\:\overline{\left|\tilde{h}_{\ell n}\right|}$. To this end, we first evaluate several quantities in $\overline{\left|\tilde{h}_{\ell m}\right|\left|\tilde{h}_{\ell n}\right|}$ and $\overline{\left|\tilde{h}_{\ell m}\right|}$. Since $h_{\ell m}$ follows the Gaussian distribution $\mathcal{N}(0,1)$, the averages of $f_{\ell m}=h_{\ell m}^2-1$, $f_{\ell m}^2$, and $|h_{\ell m}|$ are readily obtained as
\begin{align}
    \overline{f_{\ell m}}=0,\ \ \overline{f_{\ell m}^2}=2,\ \ \overline{|h_{\ell m}|}=\sqrt{\frac{2}{\pi}}.
    \label{eq:average_1}
\end{align}
Similarly, we obtain
\begin{align}
    \overline{|h_{\ell m}|f_{\ell m}}=\sqrt{\frac{2}{\pi}},\ \ 
    \overline{|h_{\ell m}|f_{\ell m}^2}=5\sqrt{\frac{2}{\pi}}.
    \label{eq:average_2}
\end{align}
In addition, the average magnitude of the nearest-neighbor hopping is 
\begin{align}
    \overline{\left|\tilde{h}_{\ell,\ell\pm1}\right|}=\sqrt{\frac{2}{\pi}},
    \label{eq:average_nearest-neighbor}
\end{align}
since $\tilde{h}_{\ell,\ell\pm1}$ is chosen to be $h_{\ell,\ell\pm1}$. For a long-range hopping between sites $\ell$ and $m$, where $\ell \neq m,m\pm1$, the multiplicative factors in Eq.~\eqref{eq:procedure_inducing-correlation} give
\begin{align}
    \overline{\left|\tilde{h}_{\ell m}\right|}
    &=\overline{\left|h_{\ell m}\right|}\prod_{s=\pm1}\overline{\left(1-\eta f_{\ell,\ell+s}\right)^2}\:\overline{\left(1-\eta f_{m,m+s}\right)^2} \prod_{n\neq\ell,\ell\pm1,m}\overline{\left(1+\eta\frac{f_{\ell,n}}{d^\kappa_{m n}}\right)^2}\prod_{n\neq\ell,m,m\pm1}\overline{\left(1+\eta\frac{f_{m,n}}{d^\kappa_{\ell n}}\right)^2}\nonumber\\
    &=\sqrt{\frac{2}{\pi}}(1+2\eta^2)^4\prod_{n\neq\ell,\ell\pm1,m}\left(1+\frac{2\eta^2}{d^{2\kappa}_{mn}}\right)\prod_{n\neq\ell,m,m\pm1}\left(1+\frac{2\eta^2}{d^{2\kappa}_{\ell n}}\right)\nonumber\\
    &\simeq\sqrt{\frac{2}{\pi}}\left[1+4\eta^2\left(2+\sum_{n\neq m}d^{-2\kappa}_{mn}\right)\right].
    \label{eq:average_long-range}
\end{align}
In the last line, we have expanded the expression up to second order in $\eta$, used Eqs.~\eqref{eq:average_1} and \eqref{eq:average_2}, and neglected terms decaying with the distance as $d_{\ell m}^{-2\kappa}$. 

We next evaluate the average product of the magnitudes of a nearest-neighbor hopping and a long-range hopping. In this case, a cross term arises between the nearest-neighbor hopping magnitude and a multiplicative factor associated with the long-range hopping. Then, for $m\neq\ell\pm1$, we obtain 
\begin{align}
    \overline{\left|\tilde{h}_{\ell m}\right|\left|\tilde{h}_{\ell \ell+1}\right|}
    &=\overline{\left|h_{\ell m}\right|}\overline{\left|h_{\ell \ell+1}\left(1-\eta f_{\ell,\ell+1}\right)^2\right|}\overline{\left(1-\eta f_{\ell,\ell-1}\right)^2}\prod_{s=\pm1}\overline{\left(1-\eta f_{m,m+s}\right)^2}\nonumber\\
    &\quad\times\prod_{n\neq\ell,\ell\pm1,m}\overline{\left(1+\eta\frac{f_{\ell,n}}{d^\kappa_{m n}}\right)^2}\prod_{n\neq\ell,m,m\pm1}\overline{\left(1+\eta\frac{f_{m,n}}{d^\kappa_{\ell n}}\right)^2}.
    \label{eq:average_product_nearest-long}
\end{align}
Evaluating Eq.~\eqref{eq:average_product_nearest-long} based on Eqs.~\eqref{eq:average_1} and \eqref{eq:average_2}, we can show
\begin{align}
    \overline{\left|\tilde{h}_{\ell m}\right|\left|\tilde{h}_{\ell \ell+1}\right|}
    &=\frac{2}{\pi}\left(1-2\eta+5\eta^2\right)(1+2\eta^2)^3\prod_{n\neq\ell,\ell\pm1,m}\left(1+\frac{2\eta^2}{d^{2\kappa}_{mn}}\right)\prod_{n\neq\ell,m,m\pm1}\left(1+\frac{2\eta^2}{d^{2\kappa}_{\ell n}}\right)\nonumber\\
    &\simeq\frac{2}{\pi}\left[1-2\eta+\eta^2\left(11+4\sum_{n\neq m}d^{-2\kappa}_{mn}\right)\right]
    \label{eq:average_product_nearest-long_approximated},
\end{align}
up to second order in $\eta$, in the same way as in Eq.~\eqref{eq:average_long-range}. Likewise, for two long-range hoppings, multiple cross terms arise, and we obtain
\begin{align}
    \overline{\left|\tilde{h}_{\ell m}\right|\left|\tilde{h}_{\ell n}\right|}
    &=\overline{\left|h_{\ell m}\left(1+\eta\frac{f_{\ell m}}{d^\kappa_{m n}}\right)^2\right|}\,\overline{\left|h_{\ell n}\left(1+\eta\frac{f_{\ell n}}{d^\kappa_{mn}}\right)^2\right|}\prod_{s=\pm1}\overline{\left(1-\eta f_{\ell,\ell+s}\right)^4}\prod_{k\neq\ell,\ell\pm1,m,n}\overline{\left(1+\eta\frac{f_{\ell,k}}{d^\kappa_{km}}\right)^2\left(1+\eta\frac{f_{\ell,k}}{d^\kappa_{kn}}\right)^2}\nonumber\\
    &\quad\times\prod_{s=\pm1}\overline{\left(1-\eta f_{m,m+s}\right)^2}\overline{\left(1-\eta f_{n,n+s}\right)^2}\overline{\left(1+\eta\frac{f_{m,n}}{d^\kappa_{\ell n}}\right)^2\left(1+\eta\frac{f_{m,n}}{d^\kappa_{\ell m}}\right)^2}\nonumber\\
    &\quad\times\prod_{k\neq m,m\pm1,\ell,n}\overline{\left(1+\eta\frac{f_{m,k}}{d^\kappa_{\ell k}}\right)^2}\prod_{k\neq n,n\pm1,\ell,m}\overline{\left(1+\eta\frac{f_{n,k}}{d^\kappa_{\ell k}}\right)^2},
        \label{eq:average_product_long-long}
\end{align}
where $m,n\neq\ell,\ell\pm1$ and $m \neq n,n\pm1$, which leads to
\begin{align}
    \overline{\left|\tilde{h}_{\ell m}\right|\left|\tilde{h}_{\ell n}\right|}
    &\simeq\frac{2}{\pi}\left(1+\frac{2\eta}{d_{mn}^\kappa}+\frac{5\eta^2}{d_{mn}^{2\kappa}}\right)^2\left(1+12\eta^2\right)^2\prod_{k\neq\ell,\ell\pm1,m,n}\left[1+\eta^2\left(\frac{8}{d^{\kappa}_{km}d^{\kappa}_{kn}}+\frac{2}{d^{2\kappa}_{km}}+\frac{2}{d^{2\kappa}_{kn}}\right)\right]\nonumber\\
    &\quad\times\left(1+2\eta^2\right)^4\left[1+\eta^2\left(\frac{8}{d_{\ell m}^{\kappa}d_{\ell n}^{\kappa}}+\frac{2}{d_{\ell m}^{2\kappa}}+\frac{2}{d_{\ell n}^{2\kappa}}\right)\right]\prod_{k\neq m,m\pm1,\ell,n}\left(1+\frac{2\eta^2}{d^{2\kappa}_{\ell k}}\right)\prod_{k\neq n,n\pm1,\ell,m}\left(1+\frac{2\eta^2}{d^{2\kappa}_{\ell k}}\right)\nonumber\\
    &\simeq\frac{2}{\pi}\left[1+\frac{4}{d_{mn}^{\kappa}}\eta+\eta^2\left(32+\frac{14}{d_{mn}^{2\kappa}}\right)\right]+\frac{2\eta^2}{\pi}\left(\sum_{k\neq\ell,\ell\pm1,m,n}\frac{8}{d^{\kappa}_{km}d^{\kappa}_{kn}}+\frac{2}{d^{2\kappa}_{km}}+\frac{2}{d^{2\kappa}_{kn}}\right)\nonumber\\
    &\quad+\frac{2\eta^2}{\pi}\left(\frac{8}{d_{\ell m}^{\kappa}d_{\ell n}^{\kappa}}+\frac{2}{d_{\ell m}^{2\kappa}}+\frac{2}{d_{\ell n}^{2\kappa}}\right)+\frac{2\eta^2}{\pi}\left(\sum_{k\neq m,m\pm1,\ell,n}\frac{2}{d^{2\kappa}_{\ell k}}+\sum_{k\neq n,n\pm1,\ell,m}\frac{2}{d^{2\kappa}_{\ell k}}\right).
    \label{eq:average_product_long-long_approximated}
\end{align}
Combining Eqs.~\eqref{eq:average_nearest-neighbor}, \eqref{eq:average_long-range}, and \eqref{eq:average_product_nearest-long_approximated}, we find that the magnitudes of the nearest-neighbor and long-range hoppings are negatively correlated,
\begin{align}
    \overline{\left|\tilde{h}_{\ell m}\right|\left|\tilde{h}_{\ell \ell+1}\right|}-\overline{\left|\tilde{h}_{\ell m}\right|}\overline{\left|\tilde{h}_{\ell \ell+1}\right|}\simeq-\frac{2}{\pi}\eta\left(2-3\eta\right),
    \label{eq:correlation_analytical_negative_appendix}
\end{align}
where $m\neq\ell,\ell\pm1$. Equation~\eqref{eq:correlation_analytical_negative_appendix} corresponds to Eq.~\eqref{eq:correlation_analytical_negative} in the main text. In contrast, the magnitudes of two long-range hoppings are positively correlated,
\begin{align}
    \overline{\left|\tilde{h}_{\ell m}\right|\left|\tilde{h}_{\ell n}\right|}-\overline{\left|\tilde{h}_{\ell m}\right|}\:\overline{\left|\tilde{h}_{\ell n}\right|}
    \simeq\frac{8}{\pi}\eta\left[d_{mn}^{-\kappa}+2\eta\left(2+\sum_{k\neq\ell,\ell\pm1,m,n}d_{km}^{-\kappa}d_{kn}^{-\kappa}\right)\right],
    \label{eq:correlation_analytical_positive_appendix}
\end{align}
where $m,n\neq\ell,\ell\pm1$ and $m \neq n,n\pm1$, which is obtained from Eqs.~\eqref{eq:average_long-range} and \eqref{eq:average_product_long-long_approximated}. Here, among the terms of order $\eta^2$, terms explicitly decaying with distances are ignored, while the constant contribution and the convolution term are retained. Equation~\eqref{eq:correlation_analytical_positive_appendix} corresponds to Eq.~\eqref{eq:correlation_analytical_positive} in the main text, with $D_\kappa=2\left(2+\sum_{k\neq\ell,\ell\pm1,m,n}d_{km}^{-\kappa}d_{kn}^{-\kappa}\right)$.
\end{widetext}

\bibliography{reference.bib}

\end{document}